# COMMISSIONING AND PERFORMANCE OF A PHASE-COMPENSATED OPTICAL LINK FOR THE AWAKE EXPERIMENT AT CERN*

D. Barrientos†, J. Molendijk, A. Butterworth, H. Damerau, W. Höfle, M. Jaussi,
CERN, Geneva, Switzerland

*Abstract*

In this work, we analyze the performance of the solution adopted for the compensation of the phase drift of a 3 km optical fiber link used for the AWAKE experiment at CERN. The link is devoted to transmit the reference signals used to synchronize the SPS beam with the experiment to have a fixed phase relation, regardless of the external conditions of the electronics and the link itself. The system has been operating for more than a year without observed drift in the beam phases. Specific measurements have proven that the jitter introduced by the system is lower than 0.6 ps and the maximum phase drift of the link is at the picosecond level.

## INTRODUCTION

The Advanced Wakefield Experiment (AWAKE) aims at studying proton-driven plasma wakefield acceleration for the first time [1]. A test facility, currently under commissioning at CERN, uses the proton beam from the SPS machine, with a momentum of 400 GeV/c, to accelerate an electron beam to the GeV scale over 10 meters of plasma. A detailed diagram of the AWAKE installation is presented in Figure 1. According to simulations, this yields an accelerating gradient of about 1 GV/m, which is more than 2 orders of magnitude larger than RF cavities currently being used.

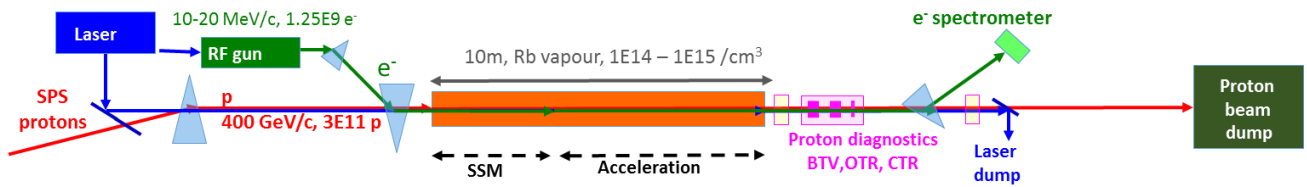

Figure 1: Baseline design of the AWAKE experiment [1] at CERN.

The LLRF system for AWAKE is synchronizing the high intensity laser pulses generating the plasma, the electron and proton beams. Laser pulses and electron beam are synchronized physically at the same place and almost no drift in their reference signals is expected, as the same laser pulse is used to drive the plasma and the electron source. However, the reference signal for the proton beam has to be transported to the SPS beam control system, about 3 km away from the location of the laser. Single-mode optical fibers used for transmitting the reference signals may introduce phase drifts due to changing environmental conditions. A precise synchronization of the beams well below the length of the proton bunch is required to get the maximum energy transfer from the proton to the electron beam and, therefore, the maximum accelerating gradient.

## PROPOSED SOLUTION

After detailed design studies, a new electronic board was developed for the synchronization of the proton beam and laser pulses in the AWAKE experiment [2-3]. In order to stabilize the phase with a precision in the picosecond level, several considerations were taken during design phase, among which are:

- Long term logging setup to evaluate maximum expected drift due to seasonal changes. In the 3 km fiber setup, the maximum expected drift is below 3 ns.
- Detailed study of phase noise contribution of SFP transceivers, introducing a jitter (integrated between 10 Hz and 10 MHz) lower than 150 fs.
- Development of the critical part of the electronics, presented in Figure 2, using differential signals to avoid common-mode noise. This part is used to transmit the 400.8 MHz RF signal (twice the frequency of the main RF system in the SPS machine) that is critical for the synchronization of the beams.
- Digitization of both polarities of the differential signal at the output of the phase detector to obtain a noise floor for the measurement that has proven to be in the 10 fs range.
- Use of delay lines with coarse and fine control. The former with 10 ps step size, allowing to cover a range of 10 ns. The latter with a range of 30 ps controlled by an analog signal generated by an external 18-bit DAC.
- Development of periodic calibration modes to compensate for non-linearity effects in delay lines, temperature changes in active devices and other effects. Each calibration mode uses a different set of delay lines by means of changing the crossbars setup in order to reject unwanted drifts in the routing of the critical path.

---

\* The authors of this work grant the arXiv.org and LLRF Workshop's International Organizing Committee a non-exclusive and irrevocable license to distribute the article, and certify that they have the right to grant this license.

† diego.barrientos@cern.ch

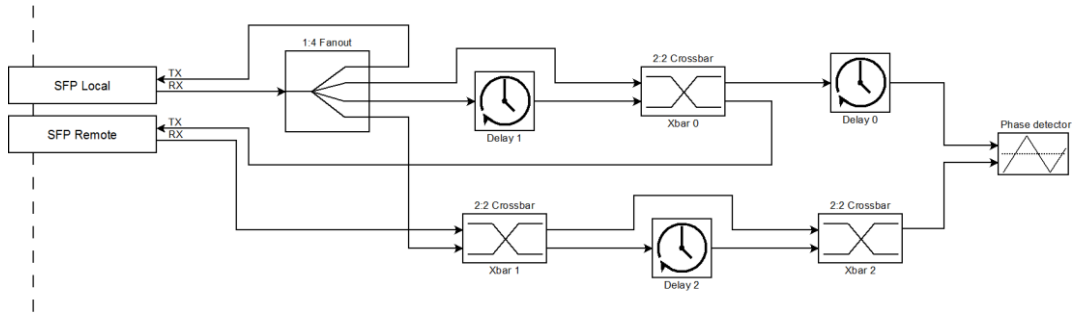

Figure 2: Block diagram of critical part of the VME board [2] devoted to measure and compensate phase drifts over the fiber link.

## HARDWARE COMMISSIONING

Following the AWAKE schedule, the first validated prototype of the module was installed during the commissioning phase of the laser and proton beam in summer 2016 [4]. Since its installation, the module has been delivering the RF reference signal used to extract the beam from the SPS machine, the common frequency between AWAKE and SPS LLRF systems and the laser repetition rate without any issue, contributing to the first physics run of AWAKE, which took place during autumn 2016.

Operational data have been recorded in a centralized logging database, as presented in Figure 3. The analysis of these data has been used to monitor the status of the system during periods without beam in the experiment. After the physics run, the AWAKE schedule foresees a commissioning phase for the electron beam, which is currently taking place. During this period, planned power cuts in the installation have shut down the system, which has shown very good reproducibility when recovering the previous state. No drifts in the relative position of the laser and proton beam have been recorded during the last year; therefore, a successful operation is expected for the next physics run at the end of the year.

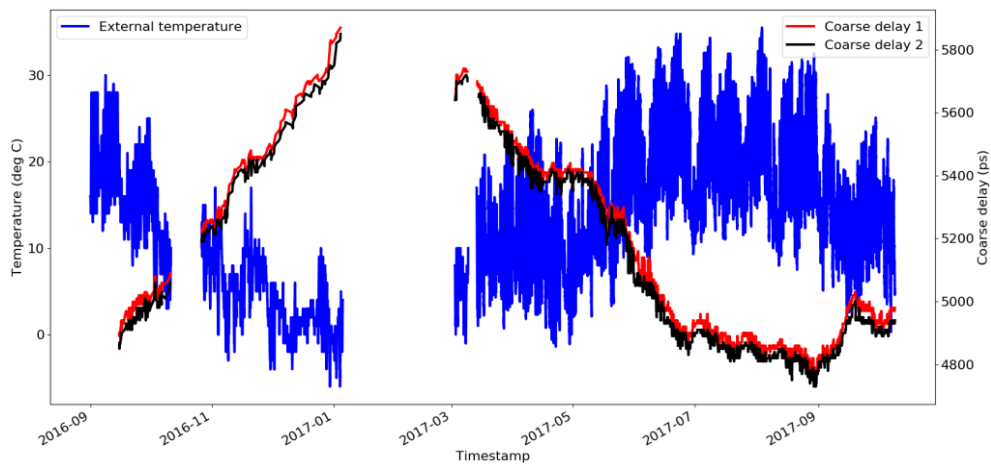

Figure 3: External temperature and coarse delay corrections.

## MODULE PERFORMANCE

In order to test the performance of the module, a test bench with two optical fibers of 3 km placed in a chamber with controlled temperature has been setup. This configuration allows accessing both ends of the long optical fiber and controlling its conditions at the same place.

A test to evaluate the added phase noise in the overall system has been performed. Firstly, the phase noise at the output of the low-noise signal generator, configured to generate a 400.8 MHz signal, is recorded. Secondly, the phase noise at the far end of the long optical fiber, after converting to an electrical signal is also recorded. Both plots, together with the noise floor curves of the instrument, are presented in Figure 4. The feedback loop compensating phase drifts in the fiber, with a bandwidth in the kHz range, was active during the test. Finally, computing the jitter from the phase noise data, integrated between 10 Hz and 10 MHz, and subtracting quadratically the jitter values. The calculated added jitter is 591 fs.

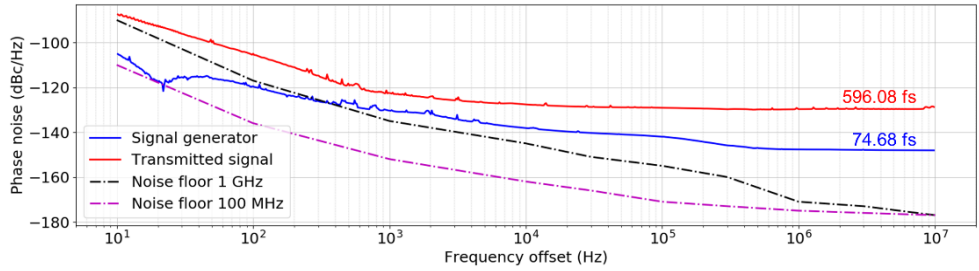

Figure 4: Phase noise plots of signal generator and transmitted signal.

Additionally, a test to measure the uncompensated phase drift when comparing signals at both ends of the long optical fiber has been carried out. An independent measurement of the phase difference has been logged using an external module. These data have been analyzed to obtain a correlation between residual phase drift and fiber temperature changes. The three tests are:

- *Optical splitter*: Setup using two long optical fibers of approximately the same length. At the far end, a copy of the signal is sent back using an optical splitter.
- *WDM*: Setup using only one long optical fiber in opposite directions by means of Wavelength-Division Multiplexing (WDM) techniques and an electrical splitter at the far end.
- *Open loop*: *WDM* setup with disabled feedback loop to evaluate phase drift without compensation.

An example of the correlation between residual phase noise and fiber temperature for the WDM test is shown in Figure 5. The results of the three tests are summarized in Table 1.

Table 1: Correlation of residual phase drift and fiber temperature changes

| Test | Phase drift |
|---|---|
| Open loop | 507.93 ps/K |
| Optical splitter | 22.72 ps/K |
| WDM | 0.61 ps/K |

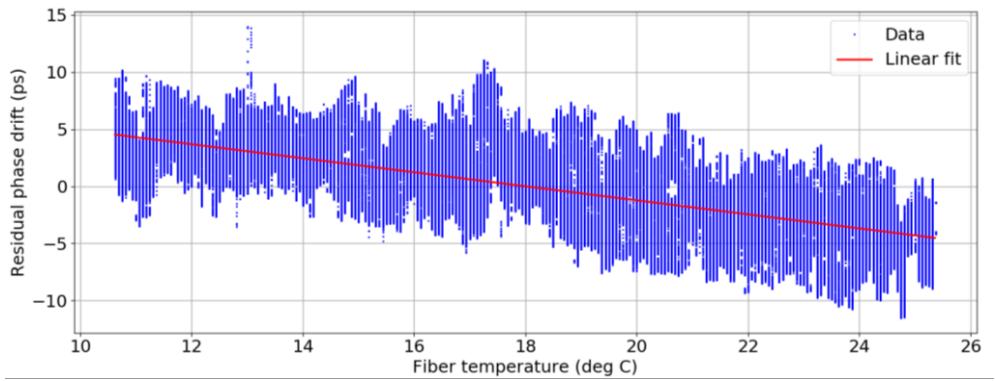

Figure 5: Phase drift vs. fiber temperature for WDM test.

## CONCLUSION

The analysis of the data recorded with the operational system in the experiment over a period of more than a year shows a maximum correction in the order of 1.2 ns in the coarse delays. Using the ratio between phase drift and temperature changes observed in the *Open loop* test in Table 1, the correction applied on the operational system corresponds to an overall temperature change in the fiber of about 2.4 K. A similar temperature span in the fiber over seasonal changes using the WDM scheme would produce a correspondent residual drift of about 1.5 ps.

In the future, long-term measurements with beam observation are planned in order to obtain the residual phase drift value of the system in the experiment conditions. Changes in the calibration procedures are also planned to be investigated in order to improve the overall performance.